\documentclass[aoas,nameyear,seceqn,rotating,dvips]{arximspdf}
\usepackage{mathbh}
\usepackage{graphicx}

\doi{10.1214/10-AOAS364}
\volume{4}
\issue{4}
\pubyear{2010}
\firstpage{1847}
\lastpage{1870}

\makeatletter

  \let\sv@tabnotetext\tabnotetext
  \let\sv@tabnotemark@fmt\tabnotemark@fmt
   \long\def\legend#1{{\let\tabnote@indent\leavevmode\sv@tabnotetext[]{}{#1}}}

\renewcommand{\citep}[1]{\citet{#1}}
\newcommand{\Ob}{\mathcal{O}}
\newcommand{\Oi}{\mathcal{O}^i}
\newcommand{\xii}{\bolds{\xi}^i}
\newcommand{\xiit}{\tilde{\bolds{\xi}^i}}
\newcommand{\btheta}{\bolds{\theta}}
\newcommand{\bbeta}{\bolds{\beta}}
\newcommand{\ui}{{\mathbf{u}}^i}
\newcommand{\bu}{\mathbf{u}}
\newcommand{\var}{\operatorname{Var }}
\newcommand{\Y}{\mathcal{Y}}
\newcommand{\sullet}{\bullet}
\makeatother

\begin{document}
\begin{frontmatter}

\title{Modeling the dynamics of biomarkers during primary HIV
infection taking
into account\break the uncertainty of infection date}
\runtitle{Modeling primary HIV infection}

\begin{aug}
\author[A]{\fnms{J.} \snm{Drylewicz}\thanksref{t1}\ead[label=e1]{julia.drylewicz@isped.u-bordeaux2.fr}},
\author[B]{\fnms{J.} \snm{Guedj}\ead[label=e2]{guedj\_jeremie@yahoo.fr}},
\author[A]{\fnms{D.} \snm{Commenges}\ead[label=e3]{daniel.commenges@isped.u-bordeaux2.fr}}
\and
\author[A]{\fnms{R.} \snm{Thi\'ebaut}\corref{}\thanksref{t2}\ead[label=e4]{rodolphe.thiebaut@isped.u-bordeaux2.fr}}

\thankstext{t1}{Received a Ph.D. grant from CASCADE Concerted Action
on SeroConversion to AIDS and Death in Europe (European Commission
Framework Program 6).}
\thankstext{t2}{On behalf of the CASCADE Collaboration.}

\runauthor{Drylewicz, Guedj, Commenges and Thi\'ebaut}
\affiliation{INSERM, U897 Epidemiology and Biostatistics and
Bordeaux 2 Victor Segalen University and Laboratory of Viral Dynamics,
Faculty of life sciences, Bar-Ilan~University 
}
\address[A]{J. Drylewicz\\
D. Commenges\\
R. Thi\'ebaut\\
INSERM, U897 Epidemiology and Biostatistics\\
Bordeaux 2 Victor Segalen University \\
146, rue Leo Saignat,
F-33076 Bordeaux\\ France\\
\printead{e1}\\
\phantom{E-mail:\ }\printead*{e3}\\
\phantom{E-mail:\ }\printead*{e4}} 
\address[B]{J. Guedj\\
Laboratory of Viral Dynamics Modeling\\
The Faculty of Life Sciences\\
Bar-Ilan University\\
52900 Ramat-Gan\\
Israel\\
\printead{e2}}
\end{aug}

\received{\smonth{5} \syear{2009}}
\revised{\smonth{5} \syear{2010}}

%
\begin{abstract}
During primary HIV infection, the kinetics of plasma virus
concentrations and CD4$+$ cell counts is very complex. Parametric
and nonparametric models have been suggested for fitting repeated
measurements of these markers. Alternatively, mechanistic
approaches based on ordinary differential equations have also been
proposed. These latter models are constructed according to
biological knowledge and take into account the complex nonlinear
interactions between viruses and cells. However, estimating the parameters
of these models is difficult. A main
difficulty in the context of primary HIV infection is that the
date of infection is generally unknown. For some patients, the
date of last negative HIV test is available in addition to the
date of first positive HIV test (seroconverters). In this paper
we propose a likelihood-based method for estimating the parameters
of dynamical models using a population approach and taking into
account the uncertainty of the infection date. We applied this
method to a sample of 761 HIV-infected patients from the
Concerted Action on SeroConversion to AIDS and Death in Europe
(CASCADE).
\end{abstract}



\end{frontmatter}

\section{Introduction}\label{s1}
Primary Human Immunodeficiency Virus (HIV) infection is a crucial
period during HIV infection history where there is a viral burst
due to the spread of the virus through target cells, mainly CD4$+$ T
lymphocytes (CD4). The dynamics of markers at that time is believed to
partly determine the evolution of the infection in the future
[\citep{mellors95}]. For instance, the peak of viral load has been
shown to be predictive of the viral setpoint, that is, the plasma HIV
RNA level at which patients often stay for several years
[\citep{Lindback00}] unless they are treated. This viral setpoint is
associated with
clinical progression [\citep{mellors95}].

Parametric and nonparametric
descriptive models have been suggested for fitting repeated
measurements of CD4 and HIV RNA (or viral load) [\citep{Dubin94}; \citep{desquilbet04}; \citep{pantazis05}; \citep{hecht06}; \citep{geskus07}]. A mechanistic approach based on ordinary
differential equations (ODE) has also been suggested
[\citep{Phillips96}; \citep{deboer98}]. These mathematical models present several
advantages. First,
they are based on biological knowledge. Therefore, the parameters may
have direct biological meaning
and the relationship between markers is modeled through biological
mechanisms rather than parametric
correlation structures. Second, this type of dynamical model is able to
capture complex
interaction between markers. For instance, these models can predict the
decrease of viral load following the peak as a
consequence of the limitation of target cells. The first models used
in this context gave important insights on the dynamics of the
infection and how to control
it [\citep{nowak1997vdp}; \citep{Little99}]. Numerous attempts to improve models have
been published [\citep{deboer98}; \citep{wick1999dct}; \citep{Stafford00}]. Most often,
the parameters of these models are not
estimated and values which appear ``reasonable'' are chosen [\citep{Phillips96}]. Indeed, estimating the parameters in such models is
difficult. To improve identifiability, random effects models (or
population approach) can be used. However, the combination
of nonlinear ODE systems and random effects leads to difficult
numerical problems [\citep{putter02}; \citep{filter2005dha};
\citep{huang2006hbm}; \citep{samson2006esa}; \citep{guedj2007mle}; \citep{cao2008epp}]: for maximizing the likelihood
we need to compute multiple integrals with a dimension equal to
the number of random effects included in the model and to solve
numerically the ODE system; all steps ask for intensive computations.

The dynamics of the biomarkers during primary HIV
infection is quite complex. In the few studies where
ODE models have been used in the context of primary infection, the
parameters were
estimated from the viral load data only
[\citep{Kaufmann98}; \citep{Little99}; \citep{Stafford00}; \citep{Lindback00}; \citep{Ciupe06}; \citep{ribeiro2010estimation}]
or the individual fit of the CD4 counts data was very bad [\citep{murray1998mph}]. Moreover, these works did not
use random effects models.

In the primary HIV infection context, the system is in a ``trivial''
equilibrium state without virus
and it is disrupted by the introduction of a small quantity of viruses
(the ``infection event'').
The date of infection must thus be known (in contrast to studies of the
response to a treatment
initiation far from the infection) if we want to compute the
trajectories of markers. However, this date is most often unknown.
Generally, the only available information is the date of the last
seronegative test and first
seropositive test or first detectable HIV RNA load. In some cohort
studies (such
as the Multicenter AIDS Cohort study [\citep{Munoz92}]) the last
seronegative date is prospectively
identified during follow-up. In most cases, patients are
seropositive at entry into the cohort and the last seronegative
date is retrospectively recorded. Usually, the date of seroconversion
is imputed
at the midpoint between last negative and first positive HIV serology test
[\citep{fidler2007scc}]. Several methods have been suggested for
estimating this date according to the marker values
[\citep{Berman90}; \citep{DeGrutolla91}; \citep{Dubin94}; \citep{Geskus00}].
\citep{geskus07} performed a conditional mean imputation to
estimate the date of infection that was used as the baseline for
linear mixed models.

In the present paper we propose a
method for estimating the parameters of dynamic models
using a population approach taking into
account the uncertainty of the infection date. We estimate
the parameters of a dynamical model using HIV RNA
and CD4 count data in a large collaboration including cohorts of
patients with a documented date of
negative and positive HIV serologies (HIV seroconverters).

There are several direct applications of such an approach. First, we
can estimate fundamental parameters such as virion clearance,
the infected cell death rate [\citep{perelson1997dch}],
virus infectivity [\citep{wilson2007eic}]
which determines the rate at which T-cells are infected for given virus
concentration,
or the basic reproduction number $R_0$ [\citep{anderson1991infectious}]
which is defined as the average number of
secondary infections that would be caused by the introduction of a
single infected cell into an entirely susceptible
population of cells; this number reflects the ability of the infection
to spread and to persist in the organism.
Second, the huge variability in the peak level of HIV RNA and of the viral
setpoint in the population [\citep{Kaufmann98}; \citep{Little99}] can be explained
by the variability of parameters that carry a direct
biological interpretation (cell death rates, virus infectivity,
virus production$,\ldots$). Third, the effect of patient
characteristics or interventions could be assessed through their
biological mechanism. For instance, one could look at the effect of an
antiretroviral regimen combining reverse transcriptase inhibitors
(blocking the infection of new cells) and protease inhibitors
(generating a higher proportion of noninfectious
virion) [\citep{Wu99}] for defining the optimal time
of treatment initiation for each patient. It would also be interesting
to look at the effect of the transmission of
mutated viruses [\citep{fidler2006cpt}] or host genetics characteristics
[\citep{fellay2007wga}] on the early dynamics.
Furthermore, the probability distribution of the date of
infection could be useful at the population level for defining the incidence
of HIV infection or the rate of disease progression and to estimate the
induction time (delay between the date
of infection and AIDS stage).

The paper is organized as follows. In Section \ref{s2} we describe a
method based on likelihood maximization for the estimation of parameters
of a general population ODE model with an unknown origin of time. In
Section \ref{s3} we
describe two HIV dynamics models and we present a simulation study.
In Section \ref{s4} we show the results for a sample of 761 HIV-infected patients
from the Concerted Action on SeroConversion to AIDS and Death in Europe
(CASCADE).
A conclusion is given in Section \ref{s5}.

\section{Statistical model and inference}\label{s2}
All the notation used in this section are summarized in Table \ref{t4} at the
end of the manuscript.
\subsection{General statistical model for systems and
observations}\label{s21}

First, we describe the model for a known date of infection. We
consider a general ODE model [two particular models are proposed in
Section \ref{s31}, equations (\ref{basic}) and (\ref{prod})];
for subject $i$ with $i = 1,\ldots,n$, we can write
\[
\cases{\displaystyle\frac{d{X^i}(t)}{dt} = f({X}^i(t),\xi^i ), \cr
X^i(0)=h(\xi^i ),}
\]
where $X^i (t)= (X^i _{1}(t),\ldots,X^i _{K}(t))$ is
the vector of the $K$ components at time $t$ and $\xi^i$ is a vector
of $p$ individual natural (or biological)
parameters which appear in the ODE system. We assume that $f$ and $h$
are known functions and twice differentiable with respect to $\xi^i$.

Second, a model is introduced for the $\xi^i$ to allow
inter-individual variability:
\[
\cases{
\tilde{\xi_{l}}^i = \Psi_l (\xi^ i_{l} ),\cr
\tilde{\xi_{l}}^i = \phi_{l}+z_{l}^ {i} \beta_{l}+ b^i,\qquad l\leq p,}
\]
where $\Psi_l$ is a known link function, $\phi_{l}$ is the intercept,
and $z^i _{l}$ is the vector
of explanatory variables associated to the fixed effects of the $l$th
biological parameter. The $\beta_{l}$'s are vectors of regression
coefficients associated to the fixed effects. We assume
$b^i \sim\mathcal{N}(0,\Sigma)$, where $b^i$ is the individual
vector of random
effects of dimension $q$. Let $A=(a_{l''l'})_{l'\leq l'' \leq q}$, the
lower triangular matrix with positive diagonal elements such that
$AA'=\Sigma$ (Cholesky decomposition). We can write $b^i=A u^i$ with
$u^i \sim\mathcal{N}(0,I_q)$.

Third, we construct a model for the observations. Generally, we do not
directly observe all
the components of $X^i$, but rather $M \leq K$ functions of $X^i$; we
call these functions
``observable components'' [see Section \ref{s32}, equations (\ref
{comp_obs1}) and (\ref{comp_obs2}), e.g.].
We observe $Y^i_{jm}$, the $j$th measurement of the $m$th
observable component for subject $i$ at date $d^i_{jm}$. If we know
the date of infection $\tau^i$, we can compute the time since
infection $t^i_{jm}=d^i_{jm}-\tau^i$ and we assume that
\[
Y^i_{jm}= g_{m}(X(t^i_{jm},\tilde{\xi}^i)) + \varepsilon^i_{jm},
\qquad j=1,\ldots,n^i_{m},\  m=1,\ldots,M,
\]
where $\tilde{\bolds{\xi}}^i=(\tilde{\xi_l}^i,l=1,\ldots,p)$ and the
$\varepsilon^i_{jm}$ are independent Gaussian measurement
errors with zero mean and variances $\sigma_{m} ^{2}$. The
$g_{m}(\cdot),m=1,\ldots,M$, are twice differentiable functions of
$\mathbb{R}^{K}$ to $\mathbb{R}$.

\subsection{The issue of unknown date of infection}\label{s22}

We assume that the date of infection $\tau^{i}$ has a known
density $f_{\tau^i}$ with a support $[\tau_{\inf}^i,\tau_{\sup}^i]$. We
denote by $d_-^i$ (resp. $d_+^i$) the date of last seronegative
test (resp. first seropositive test) for patient $i$.
For very long intervals it would be relevant to
take the density proportional to HIV incidence. Since we work with
intervals of moderate length (3 years maximum), the incidence can be
considered approximately constant. Therefore, we took uniform densities
between $d_-^i$ and $\tau_{\sup}^i=\min(d_+^i,d^i_{11})$, where $d^i_{11}$
is the date of first HIV RNA detectable measurement.

However, another problem is that patients may have been infected before
the last seronegative date. \citep{fiebig2003dhv}
determined that the HIV serology becomes positive on average 89 days
(95\% Confidence Interval${}={}$47--130) after infection. The window of
possible infection is thus $[\tau_{\inf}^i,\tau_{\sup}^i]$, where $\tau
_{\inf}^i=(d_-^i-130)$.
However, it is less likely that a patient has been infected 130 days
before $d_-^i$ than 5 days before $d_-^i$, for instance. Therefore, we assumed
linearly increasing densities between $\tau_{\inf}^i$ and $d_-^i$ starting
with $f_{\tau^i}(\tau_{\inf}^i)=0$. Thus, the density $f_{\tau^i}$ is
defined by
%
\begin{equation}\label{density}
f_{\tau^i}(s)= \cases{
0,&\quad\mbox{if }$s \le\tau_{\inf}^i$,\cr
\displaystyle\frac{f_{\tau^i}(d_-^i)(s-\tau_{\inf}^i)}{d_-^i-\tau_{\inf}^i},&\quad\mbox{if }$\tau_{\inf}^i \le s \le
d_-^i$,\cr
f_{\tau^i}(d_-^i),&\quad\mbox{if }$s \ge d_-^i$,}
\end{equation}
where $f_{\tau^i}(d_-^i)=2/(2\tau_{\sup}^i-d_-^i-\tau
_{\inf}^i)$ to ensure the continuity of the density
and\ $\int f_{\tau^i}(s)\,ds=1$.

\subsection{Log-likelihood}

The model is complicated by the detection limits $\zeta^i_{j}$ of
assays leading to left-censored observations of HIV RNA for the
$j$th measurement for subject $i$. We define HIV RNA as the first
observable compartment ($m=1$) and we observe $Y^i_{j1}$ or
$\{Y^i_{j1} < \zeta^i_{j}\}$. The left-censored observations are
taken into account as in \citep{thiebaut2006edm} and
\citep{guedj2007mle}. Noting $\delta_{ij}=\mathbh{1}_{\{Y^i_{j1} >
\zeta^i_{j}\}}$, 
the full likelihood given the random effects $u^i$ and the date of
infection $\tau^{i}$ is given by
\begin{eqnarray*}
\mathcal{L}_{\mathcal{Y}^i |{u^i,\tau^{i}}} &=&
\prod^{n^i_{1}}_{j=1} \biggl\{ \frac{1}{\sigma_{1}\sqrt{2\pi
}} \exp\biggl[
-\frac{1}{2}\biggl( \frac{ Y^i_{j1} - g_{1}(X(t^i_{j1},\xiit))} {\sigma_{1}}\biggr)^{2}\biggr]
\biggr\}^{\delta_{ij}}\\
&&\hspace*{13pt}{}\times
\biggl\{ \Phi\biggl(\frac{\zeta^i_{j} - g_{1}(X(t^i_{j1},\xiit))}{\sigma_{1}} \biggr)
\biggr\}^{1-\delta_{ij}} \\
&&{}\times
\mathop{\prod_{m=2,M}}_{j=1,n^i_{m}}\biggl\{ \frac{1}{\sigma_{m}\sqrt{2\pi}} \exp\biggl[-
\frac{1}{2}\biggl(\frac{ Y^i_{jm} - g_{m}(X(t^i_{jm},\xiit))} {\sigma
_{m}}\biggr)^{2} \biggr]
\biggr\},
\end{eqnarray*}
where $\Phi$ is the cumulative distribution function of the
standard univariate normal distribution. The individual likelihood
given the date of infection is
\[
\mathcal{L}_{\Oi|\tau^{i}} = \int_{\mathbb{R}^{q}}
\mathcal{L}_{\mathcal{Y}^i |{u^i,\tau^{i}}}(u) \phi(u)\,du,
\]
where $\phi$ is the multivariate normal density of $\mathcal{N}(0,I_{q})$.
To determine the observed individual likelihood
($\mathcal{L}_{\Oi}$), we integrate $\mathcal{L}_{\Oi|\tau^{i}}$
on $[\tau_{\inf}^i;\tau_{\sup}^i]$:
\[
\mathcal{L}_{\Oi} = \int_{\tau_{\inf}^i}^{\tau_{\sup}^i}
\mathcal{L}_{\Oi|\tau^{i}}(s)f_{\tau^i}(s)\,ds =
\int_{\tau_{\inf}^i}^{\tau_{\sup}^i}\int_{\mathbb{R}^{q}}
\mathcal{L}_{ \mathcal{Y}^i |{u^i,\tau^{i}} }(u,s)
\phi(u) f_{\tau^i}(s)\,du\,ds.
\]
We note $L_{\mathcal{Y}^i |{u^i,\tau^{i}}}=\log\mathcal
{L}_{\mathcal{Y}^i
|{u^i,\tau^{i}}}$ and $L_{\Oi}=\log \mathcal{L}_{\Oi}$.
The global
observed log-likelihood is $L_{\mathcal{O}}=\sum_{i \le n}L_{\Oi}$.
The integral on time is calculated by recursive adaptive Gauss-Legendre
quadrature [\citep{berlizov1999recursive}].

\subsection{Likelihood maximization}

A Newton-like method was used for maximizing the likelihood. This method
uses the first derivatives of the log-likelihood (the scores)
[\citep{Commenges06}]. The observed scores
can be obtained by applying twice Louis' formula [\citep{Louis82}]:
\[
U_{\Ob^i} = \frac{ \partial L_{\Ob^i}}{\partial\theta} = (\mathcal
{L}_{\Oi}) ^{-1}
\int_{\tau_{\inf}^i}^{\tau_{\sup}^i} \mathcal{L}_{\Oi|\tau^{i}}(s)
U_{\Oi|\tau^{i}}
(s)f_{\tau^i}(s)\,ds ,
\]
where:
\[
U_{\Oi|\tau^{i}}(\cdot) = (\mathcal{L}_{\Oi|\tau^{i}}(\cdot)) ^{-1} \int
_{\mathbb{R}^{q}}
\mathcal{L}_{\mathcal{Y}^i |{u^i,\tau^{i}}}(u,\cdot)
U_{\mathcal{Y}^i |{u^i,\tau^{i}}} (u) \phi(u)\, du.
\]
The scores $U_{\mathcal{Y}^i |{u^i,\tau^{i}}}$ are determined by
differentiating $L_{\mathcal{Y}^i |{u^i,\tau^{i}}}$ by $\theta$,
where $\theta=((\phi_l)_{l=1,p},(\beta_l)_{l=1,p},A=(a_{ll'})_{l'
\leq l \leq q},\sigma=(\sigma_l)_{l \le M} )$.

\subsection{A posteriori estimations and distribution of the
date of infection}

Individual parameters $\tilde{\xi} ^i$ were estimated by the
empirical Bayes estimators $\hat{\tilde{\xi}^i}$, where $\hat{
\tilde{\xi}
}_l^i=\hat{ \phi}_l + z^{i}_l \hat{ \beta}_l + \hat{A}
\hat{u}^i$ and $\hat{u}^i$ is the posterior mode of the density of
$u^i$ given the data. Then, individual trajectories were predicted by
computing $\hat{X}^i=X(t,\hat{ \tilde{\xi} ^i})$.

An estimator of the a posteriori distribution of the date of
infection, $f_{\tau^i|Y^i}$, can then be computed at the estimated individual
parameters $\hat{\tilde{\xi}^i}$. We denote by $Y^i$ the vector of
measurements for
subject $i$ and by $f_{Y^i|\tau^i}$ the probability density
function of $Y^i$ given the date of infection. Applying Bayes'
formula, we have
\[
f_{ \tau^i|Y^i }(\cdot|Y^i)=\frac{ f_{Y^i|\tau^i}(Y^i|\cdot) f_{\tau^i}(\cdot) }
{ \int_{\tau_{\inf}^i}^{\tau_{\sup}^i} f_{Y^i|\tau^i} (Y^i|s) f_{\tau^i}(s)
\,ds }.
\]

\section{Models for primary HIV infection}\label{s3}
\subsection{Models for the biological system}\label{s31}

We considered two ODE models for HIV infection: the ``basic model''
and\vadjust{\goodbreak}
the ``productive cells model.''
The ``basic model'' has three compartments:
$T$ (noninfected CD4), $T^*$ (productively infected CD4) and
$V$ (free virion) [\citep{nowak1997vdp}; \citep{Stafford00}; \citep{perelson2002mva}].
This model can be written as
%
\begin{equation}\label{basic}
\cases{
\displaystyle\frac{dT}{dt}= \lambda- \gamma VT - \mu_T T,\vspace*{2pt} \cr
\displaystyle\frac{dT^*}{dt}= \gamma VT -\mu_{T^*} T^*,\vspace*{2pt} \cr
\displaystyle\frac{dV}{dt}= \pi T^* - \mu_V V.
}
\end{equation}

\begin{figure}

\includegraphics{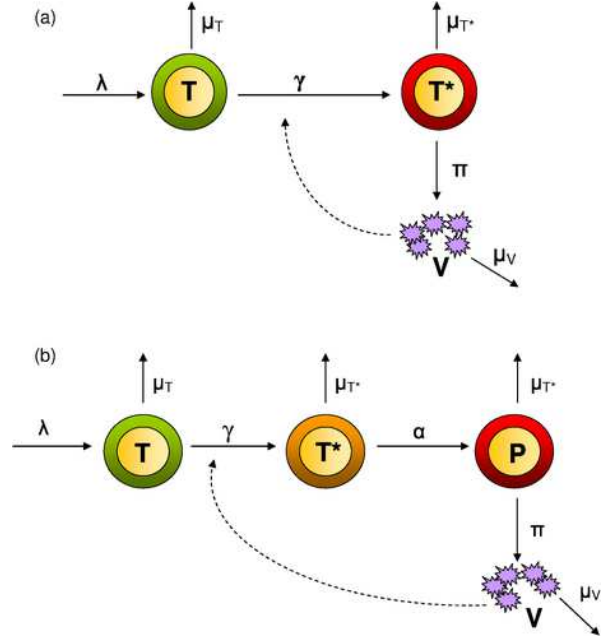}

\caption{Graphical representation of ``the basic model'' \textup{(a)} and ``the
productive cells model''~\textup{(b)}.}\label{model}
\end{figure}

The uninfected CD4 cells enter the blood circulation at rate $\lambda$,
and die at the rate $\mu_T$. They can be infected by the virus at the rate
$\gamma V$. The infected CD4 die at the rate
$\mu_{T^*}$ and produce virions at the rate $\pi$.
Virions die at the rate $\mu_V$ and can also infect other CD4 cells
[Figure \ref{model} (a)]. Model parameters are summarized in
Table~\ref{param}.

\begin{table}
\caption{Parameters of the dynamic models}\label{param}
\begin{tabular*}{\textwidth}{@{\extracolsep{\fill}}lll@{}}
\hline
\textbf{Parameter} &  \multicolumn{1}{l}{\textbf{Meaning (per day)}} &  \multicolumn{1}{l@{}}{\textbf{Value}}\\
\hline
$\lambda$ &Rate of $T$ cell production per $\mu$L & estimated\\
$\mu_{T^*}$ &Death rate of $T^*$ cells & estimated\\
$\pi$ &Number of virions per $T^*$ cells & estimated\\
$\mu_{T}$ &Death rate of $T$ cells & estimated\\
$\gamma$ &Infection rate of $T$ cells per virion per $\mu$L &
0.00027 [\citep{davenport2006ipv};\\
&& \citep{wilson2007eic}]\\
$\mu_V$ &Clearance of free virions & 20.0 [\citep{Ramratnam99}]\\
$\alpha$&Rate to become productive cells & 1.0 [\citep{kiernan1990khr};\\
&& \citep{barbosa1994kah}]\\
\hline
\end{tabular*}
\end{table}

We assume that the model is at equilibrium before infection, so at
$t=0_-$, we have
\[
\cases{
T_{0_-} =\displaystyle\frac{\lambda}{\mu_T}, \cr
T^*_{0_-} = 0, \cr
V_{0_-} = 0.}
\]

The initial inoculum of virions is fixed at $10^{-6}$ virions/mm$^3$
[\citep{Ciupe06} \citep{Stafford00}], which
is equivalent to 5 virions for 5 liters of blood. The introduction
of virions in the system at $t=0$ disrupts the initial stability and
the system stabilizes to a new
equilibrium if the basic reproduction number $R_0$ ($R_0=\frac{\lambda
\gamma\pi}{\mu_T\mu_{T^*}\mu_V}$) is higher than 1:
%
\[
\cases{\bar{T} = \displaystyle\frac{\mu_{T^*}\mu_V}{\gamma\pi},\vspace*{2pt} \cr
\bar{T}^* = \displaystyle\frac{\lambda\gamma\pi-\mu_T\mu_{T^*}\mu_V}{\gamma
\pi\mu_{T^*}},\vspace*{2pt} \cr
\bar{V} = \displaystyle\frac{\lambda\gamma\pi-\mu_T\mu_{T^*}\mu_V}{\gamma
\mu_{T^*}\mu_V}.
}
\]

The second model (the ``productive cells model'') distinguishes
nonproductive infected cells ($T^*$)
and productive infected cells ($P$) [Figure \ref{model}(b)] as previously
suggested in \citep{nowak1997avd}. This second model includes
one more parameter $\alpha$ (the rate to become productive infected cells).
The model can be written as
\begin{equation} \label{prod}
\cases{
\displaystyle\frac{dT}{dt}= \lambda- \gamma VT - \mu_T T,\vspace*{2pt} \cr
\displaystyle\frac{dT^*}{dt}= \gamma VT -\mu_{T^*} T^* -\alpha T^*,\vspace*{2pt}\cr
\displaystyle\frac{dP}{dt}= \alpha T^* -\mu_{T^*} P,\vspace*{2pt}\cr
\displaystyle\frac{dV}{dt}= \pi P - \mu_V V.}
\end{equation}

The equilibrium before infection is
\[
\cases{
T_{0_-} = \displaystyle\frac{\lambda}{\mu_T} ,\vspace*{2pt}\cr
T^*_{0_-} = 0 ,\cr
P_{0_-} =0 ,\cr
V_{0_-} = 0.}
\]
After the introduction of virions, the system
stabilizes to a new equilibrium if $R_0$ ($R_0=\frac{\lambda\gamma
\pi\alpha}{\mu_T\mu_{T^*}\mu_V(\mu_{T^*}+\alpha)}$)
is higher than 1:
\[
\cases{
\displaystyle\bar{T} = \frac{\mu_{T^*}\mu_V(\mu_{T^*}+\alpha)}{\gamma\pi\alpha}, \cr
\displaystyle\bar{T}^* = \frac{\lambda}{\mu_{T^*}+\alpha}-\frac{\mu_T\mu_V\mu_{T^*}}{\gamma\pi\alpha} ,\vspace*{2pt}\cr
\displaystyle\bar{P} = \frac{\alpha}{\mu_{T^*}}\biggl(\frac{\lambda}{\mu_{T^*}+\alpha}-\frac{\mu_T\mu_V\mu_{T^*}}{\gamma\pi\alpha}\biggr),\vspace*{2pt}\cr
\displaystyle\bar{V} = \frac{\pi\alpha}{\mu_V\mu_{T^*}}\biggl(\frac{\lambda}{\mu_{T^*}+\alpha}-\frac{\mu_T\mu_V\mu_{T^*}}{\gamma\pi\alpha}\biggr).}
\]

\subsection{Statistical models}\label{s32}

We can construct statistical models as in Section \ref{s21}. We took $\Psi
_l(\cdot)=\log(\cdot)$ for all $l$ to ensure positivity of parameters.
The individual parameters are $\tilde{\bolds{\xi}}^i=(\tilde{\lambda}^{i}, \tilde{\mu}_{T^*}^i, \tilde{\pi}^i, \tilde{\mu}_T^i)$ with
$\tilde\lambda^i=\log\lambda^i$, $\tilde\mu_{T^*}^i=\log\mu
_{T^*}^i$, $\tilde\pi^i=\log\pi^i$ and $\tilde\mu_T^i=\log\mu_T^i$.
Because of identifiability issues [\citep{wu2008parameter}],
the values of the other parameters were taken according to the literature:
$\mu_V=20.0$ day$^{-1}$ [\citep{Ramratnam99}] and $\gamma =0.00027$
virions$^{-1}$ day$^{-1}$ $\mu$L$^{-1}$
[\citep{davenport2006ipv}; \citep{wilson2007eic}]. The parameter $\alpha$ is
fixed at 1 day$^{-1}$ because the time lag between virus
entry and virus production is \mbox{about 1 day} [\citep{kiernan1990khr}; \citep{barbosa1994kah}].

The observed components were the base-10 logarithm of HIV RNA load
(number of virions per $\mu$L) and the fourth root of total CD4 count
(number of cells per $\mu$L).
For the ``basic model,'' $g_1(X)=\log_{10}(V)$ and
$g_2(X)=(T+T^*)^{0.25}$ where $X=(T,T^*,V)$.
For the ``productive cells model,'' $g_1(X)=\log_{10}(V)$ and
$g_2(X)=(T+T^*+P)^{0.25}$ where $X=(T,T^*,P,V)$.
These transformations are commonly used
for achieving normality and homoscedasticity of measurement error
distributions [\citep{Thiebaut03}]. We have
%
\begin{eqnarray}
Y^i_{j1} &=&g_{1}(X(t^i_{j1},\tilde{\xi}^i)) + \varepsilon^i_{j1},\qquad
j=1,\ldots,n^i_{1},\label{comp_obs1}\\
Y^i_{j2} &=&g_{2}(X(t^i_{j2},\tilde{\xi}^i)) +
\varepsilon^i_{j2},\qquad
j=1,\ldots,n^i_{2}\label{comp_obs2},
\end{eqnarray}
where $\varepsilon^i_{j1}$ and $\varepsilon^i_{j2}$ are independent
Gaussian with zero mean and variances
$\sigma_{\mathit{VL}} ^{2}$ and $\sigma_{\mathit{CD}4} ^{2}$ respectively.

\subsection{Simulation study}\label{s33}

We simulated samples of 100 subjects during primary HIV
infection using the ``basic model'' for simplicity. The parameter values
were defined according to the estimates of our application (see Tables
\ref{param} and \ref{para_est}).
For each subject, we simulated the dates of HIV tests and the date of
infection. Dates for the HIV serology tests (negative and positive,
respectively) were simulated
according to the prior distribution defined in Section \ref{s21} [equation
(\ref{density})]. The subjects had a probability
of 0.10 to have a short window of infection (90 days) with repeated
measurements around the time of viral peak and a
probability of 0.90 to have a window of infection of 200 days with
repeated measurements after 150 days post-infection. The
times of measurements $t^i_{jm}$ were as follows: days 1, 10, 15, 30,
50 and 100
for the short windows and days 150, 200, 240, 280, 300 and 350 for the others.
We included independent random effects for the first two parameters
($\lambda$ and $\mu_{T^*}$).
Therefore, the vector of parameters to be estimated was $(\tilde
\lambda^i, \tilde\mu_{T^*}^i, \tilde\pi^i, \tilde\mu_T^i, \sigma_{\tilde{\lambda}},
\sigma_{\tilde{\mu}_{T^*}}, \sigma_{\mathit{VL}}, \sigma_{\mathit{CD}4})$.

We performed a simulation study to compare the estimates according to
three methods: (i) when the date of infection is fixed as the real date
(RD) or (ii) when it is imputed at the midpoint of the interval defined
the last negative and the first positive HIV test (DI) and
(iii) when the uncertainty of the date of infection was taken into
account with our proposed method (DUK). We simulated 50 data sets of 100
subjects with the design described above. For each data set, we estimated
parameters with the three methods and we compared them by computing the
mean square errors and the coverage rates (Table \ref{simu_comp}). The
method using
the real date of infection was used as a reference.

\begin{table}[b]
\caption{Mean Square Errors (MSE) and coverage rates of the estimation
method in the situations where the date of infection is
known (RD), the date of infection is imputed at the midpoint (DI) and
the uncertainty of the date of infection is
taken into account (DUK)}
\label{simu_comp}
\begin{tabular*}{\textwidth}{@{\extracolsep{\fill}}lcccrrr@{}}
\hline
 & \multicolumn{3}{c}{\textbf{MSE}}  &\multicolumn{3}{c@{}}{\textbf{Coverage rate ($\bolds{\%}$)}}\\
 &\multicolumn{3}{c}{\hrulefill}&\multicolumn{3}{c@{}}{\hrulefill}\\
 \textbf{Parameter}  & \multicolumn{1}{c}{\textbf{RD}} & \multicolumn{1}{c}{\textbf{DI}} & \multicolumn{1}{c}{\textbf{DUK}} & \multicolumn{1}{c}{\textbf{RD}} & \multicolumn{1}{c}{\textbf{DI}} & \multicolumn{1}{c@{}}{\textbf{DUK}} \\
\hline
$\tilde{\lambda}$ & 0.0030 & 0.0498 & 0.0033 & 100 &80 & 100 \\
$\tilde{\mu}_{T^*}$ & 0.0098 & 0.1880 & 0.0086 & 100 &90 & 100 \\
$\tilde{\pi}$ & 0.0021 & 0.0426 & 0.0115 & 100 &90 & 98 \\
$\tilde{\mu}_{T}$& 0.0065 & 4.0268 & 0.0023 & 100 &60 & 100 \\

$\sigma_{\tilde{\lambda}}$ & 0.0014 & 0.0079 & 0.0022 & 100 &75 &100 \\
$\sigma_{\tilde{\mu}_{T^*}}$ & 0.0051 & 0.0143 & 0.0071 & 100 &90& 98 \\
$\sigma_{\mathit{VL}}$ & 0.0011 & 0.0075 & 0.0007 & 100 & 100 & 100 \\
$\sigma_{\mathit{CD}4}$ & 0.0001 & 0.0688 & 0.0001 & 100 & 100 & 100 \\
\hline
\end{tabular*}
\legend{$\sigma_{\tilde{\lambda}}$ and $\sigma_{\tilde{\mu}_{T^*}}$ are
the standard deviations
of random effect for $\tilde{\lambda}$ and $\tilde{\mu}_{T^*}$.\\
$\sigma_{\mathit{VL}}$ and $\sigma_{\mathit{CD}4}$ are the standard deviations of the
observation error
of $\log_{10}(V)$ and $(T+T^*)^{0.25}$.}
\end{table}

The DUK estimators had a much smaller MSE than the DI estimators and
close to the MSE of the RD estimators. It is not feasible with such
computationally demanding
programs to make a large simulation. Fifty replications are not enough
to precisely estimate coverage rates but are enough to show the
superiority of our approach
over the approach based on imputing the date of infection at midpoint.
The empirical coverage rates were slightly too high for RD and DUK
probably due to overestimation of the variance of the parameters.

\section{Application to the CASCADE data set}\label{s4}
\subsection{The study sample}
The study sample came from the Concerted Action on
SeroConversion to AIDS and Death in Europe (CASCADE) study; it includes
seroconverters
from Europe, Canada and Australia. The
CASCADE study has been described in detail elsewhere
[\citep{cascade2003dsf}]. Data were pooled in the 2006 update from
20 participating cohorts. We selected HIV seroconverters if their
HIV interval test (delay between the date of last seronegative test
and the date of first seropositive test) was less than 3 years, if
they did not receive any antiretroviral treatment during the first
year following the date of seropositive test and if they had
more than 3 measurements of CD4 or viral load during this first year of
follow-up with
the first detectable viral load measurement during the first three
months after the
date of seropositive test. A total of 761 patients met these criteria.

\subsection{Models used}

We used the two models defined in Section \ref{s31}.
The vector of natural parameters to be estimated for subject $i$ was
$\xi^i=(\lambda^i, \mu_{T^*}^i,\break \pi^i, \mu_T^i)$ for the two models.
The other parameters were assumed to be known as described in Section
\ref{s32}.
The windows for possible dates of infection were fixed as defined in
Section \ref{s22}.

Random effects were selected
according to a forward selection procedure. Starting with a model without
random effect, we introduced random effects successively on each
parameter and selected the one leading to the best likelihood
improvement. Then, we added a new random effect and continued
until the new model was not rejected by a likelihood ratio test.

\subsection{Results}

Selected patients had a median of four measurements for CD4 and for HIV RNA
(InterQuartile Range [IQR$]= [$3; 5]). Most of the patients were infected
by sex between men (71\%).
Follow-up was censored after 1 year beyond seropositive HIV test,
resulting in a median follow-up after the
first seropositive test of 195 days (IQR${}= [$119; 260]). The median of
the delay between the dates of last seronegative
and first seropositive test was 170 days (IQR${}= [$91; 273]).

The final models included independent random effects for the first two
parameters. Therefore, the estimated vector of parameters was
$(\tilde{\lambda}, \tilde{\mu_{T^*}}, \tilde{\pi}, \tilde{\mu
_T}, \sigma_{\tilde{\lambda}},\break \sigma_{\tilde{\mu}_{T^*}}, \sigma
_{\mathit{VL}}, \sigma_{\mathit{CD}4})$
for the two models. Usually, random effects are assumed to be
independent in ODE models [\citep{putter02}; \citep{samson2006esa}; \citep{huang2006bayesian}; \citep{guedj2007mle}]. To
test a possible correlation between the two random effects included in
our models, we developed a score test based on our previous work [\citep{Drylewicz2010scoretest}].
For both models, the test was not significant ($p=0.87$ and $p=0.92$,
respectively). The score test is presented in Appendix \ref{apenA}.

The ``productive cells model'' fitted better than the ``basic model''
(AIC${}={}$ 14,300 vs. 15,010). The improvement brought by this model can be
considered as large according to the criterion $\mathrm{D}$ which is
the difference of AIC divided by the total number of observations:
$\mathrm{D}=\frac{15{,}010-14{,}300}{6294}=0.11$ [\citep{commenges2008edk}].
Estimates of the parameters of the two models are presented in Table~\ref{para_est}.

\begin{table}
\tablewidth=300pt
\caption{Estimated parameters and standard-errors (SE) on logarithmic
scale for the ``basic''
and the ``productive cells'' models; for meaning of parameters see
Table \protect\ref{param}}\label{para_est}
\begin{tabular*}{300pt}{@{\extracolsep{\fill}}lcccccc@{}}
\hline
&\multicolumn{3}{c}{\textbf{Basic model}} & \multicolumn{3}{c@{}}{\textbf{Productive cells model}}\\
&\multicolumn{3}{c}{\hrulefill}&\multicolumn{3}{c@{}}{\hrulefill}\\
\textbf{Parameter}&\multicolumn{1}{c}{\textbf{ Estimate}}&&\multicolumn{1}{c}{\textbf{SE}}&\multicolumn{1}{c}{\textbf{Estimate}}&&\multicolumn{1}{c@{}}{\textbf{SE}}\\
\hline
$\tilde{\lambda}$&\phantom{$-$}3.49&&0.032&\phantom{$-$}3.33&&0.034\\
$\tilde{\mu_{T^*}}$&\phantom{$-$}0.62&&0.046&\phantom{$-$}0.54&&0.045\\
$\tilde{\pi}$&\phantom{$-$}5.27&&0.041&\phantom{$-$}6.06&&0.036\\
$\tilde{\mu_T}$&$-$3.26 &&0.019&$-$3.40&&0.034\\
$\sigma_{\tilde{\lambda}}$&\phantom{$-$}0.16&&0.008&\phantom{$-$}0.41&&0.057\\
$\sigma_{\tilde{\mu}_{T^*}}$&\phantom{$-$}0.42&&0.043&\phantom{$-$}0.37&&0.052\\
$\sigma_{\mathit{VL}}$&\phantom{$-$}0.82 &&0.017&\phantom{$-$}0.75&&0.017\\
$\sigma_{\mathit{CD}4}$&\phantom{$-$}0.29&&0.012&\phantom{$-$}0.27&&0.013\\[5pt]
AIC&&15,010&&&14,300&\\
\hline
\end{tabular*}
\legend{$\sigma_{\tilde{\lambda}}$ and $\sigma_{\tilde{\mu}_{T^*}}$
are the standard deviations of random effect for $\tilde{\lambda}$
and $\tilde{\mu}_{T^*}$.\break
$\sigma_{\mathit{VL}}$ and $\sigma_{\mathit{CD}4}$ are the standard deviations of the
observation
error of $\log_{10}(V)$ and $(T+T^*)^{0.25}$ for the ``basic model''
and $(T+T^*+P)^{0.25}$ for the ``productive cells model.''}
\end{table}

For the ``productive cells model,'' on the natural scale,
the mean half-life [$\log(2)/$Death rate] of infected and
uninfected cells was 0.40 and 21 days,
respectively. At the time of the viral peak (median viral load of 5.68
$\log_{10}$ copies/mL, $\mathrm{IQR}= [5.44; 5.96]$),
the estimated median number of infected cells was 64~cells/$\mu$L
($\mathrm{IQR}= [40; 113]$). At the same time, the median number of
productive cells/$\mu$L was 24 ($\mathrm{IQR}= [14; 46]$) among 557 CD4
cells/$\mu$L
($\mathrm{IQR}= [460; 658]$). The median of the estimated CD4 setpoints was 500
cells/$\mu$L ($\mathrm{IQR}= [396; 627]$)
which is close to the median of observed setpoints (based on
at least one measurement after 365 days available in 430 patients): 470
cells/$\mu$L ($\mathrm{IQR}= [371; 615]$). At the same time,
the median of the estimated number of infected cells/$\mu$L was 8
($\mathrm{IQR}= [6; 12]$) including a median of 3 productive
cells/$\mu$L ($\mathrm{IQR}= [2; 5]$). The median of viral setpoints was also
close to that observed in 403 patients:
\mbox{4.77 $\log_{10}$ copies/mL} ($\mathrm{IQR}= [4.59; 4.98]$) vs. 4.59 ($\mathrm{IQR}= [4.04; 5.00]$).

\begin{figure}

\includegraphics{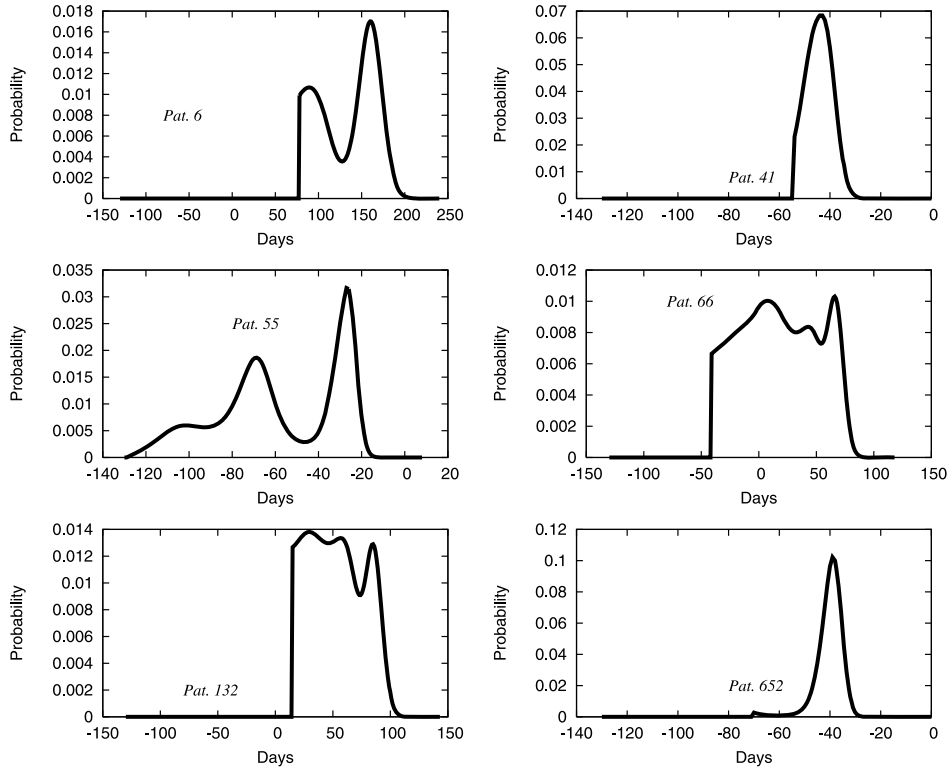}

\caption{A posteriori individual distributions of the date
of infection for patients 6, 41, 55, 66, 132 and 652 from the
``productive cells model,''
where day 0 is the date of last negative HIV test.}\label{dist_ind}
\end{figure}

\begin{figure}

\includegraphics{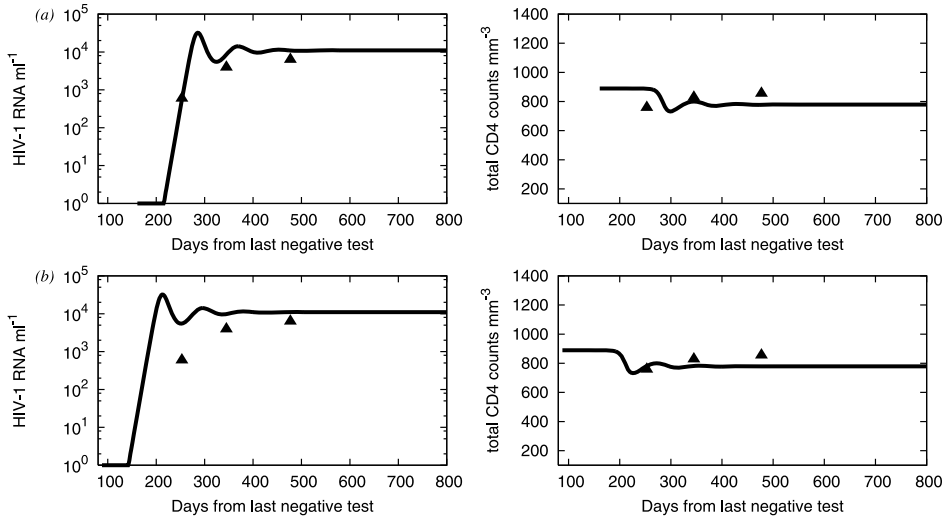}

\caption{Predicted fits from the two most probable dates of infection
[\textup{(a)} the global maximum and \textup{(b)} the local maximum] and observed values of
HIV RNA level and total CD4 count for patient 6 from the ``productive
cells model.''}\label{locmax}
\end{figure}

We found very different a posteriori individual distributions of
the date of infection depending on the width of the window and the quantity
of available information for each patient. Figure \ref{dist_ind} shows the
a posteriori distributions displayed on the window of infection
(see Section \ref{s22}), where 0 represents the last seronegative date for six
patients chosen for exhibiting different shapes of the a posteriori
distribution. For patients with data clearly before the setpoint, the
model was
able to restrict the possible dates of infection. For instance, patient 132
had an interval of possible dates of infection of 273 days and our
estimation restricted this interval to
90 days. However, for other patients and especially for some patients
with a wide interval between last negative and first positive HIV
serology test,
the a posteriori distribution had
local maxima. Analysis of marker dynamics according to the local maxima
suggested quite different possibilities according to the available
information (Figure \ref{locmax}).

\begin{figure}

\includegraphics{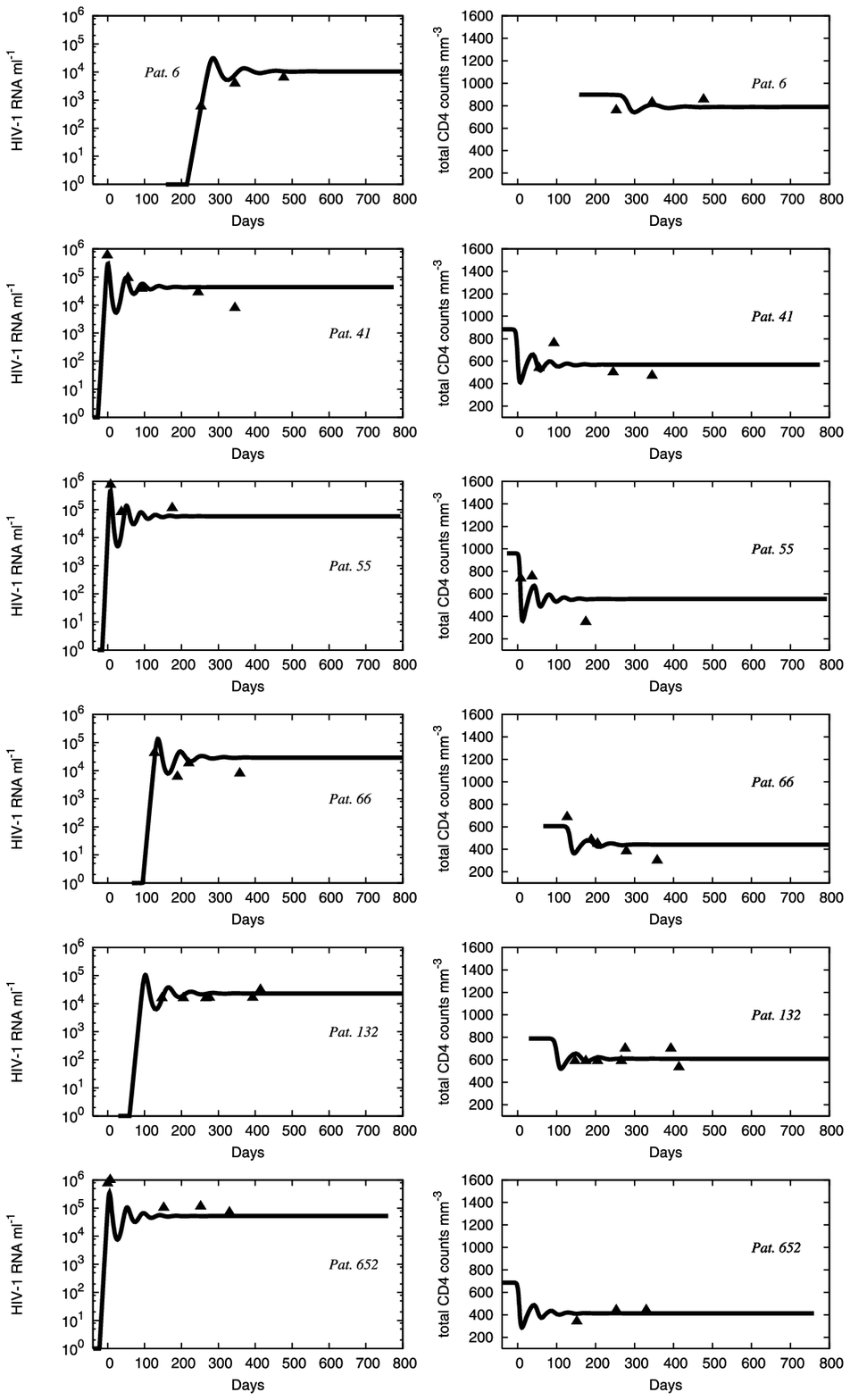}

\caption{Individual predicted
fits from the most probable date of infection and observed values
of HIV RNA level and total CD4 count for patients 6, 41, 55, 66, 132
and 652 from the ``productive cells model,''
where day 0 is the date of last negative HIV test.}\label{fit_ind}
\end{figure}

For each patient, we took the date with the highest probability
and plotted predicted individual trajectories for each marker
from this date. The fits of the model were satisfactory. The estimated
trajectories are in agreement with those reported in the literature
[\citep{Little99}; \citep{Lindback00}; \citep{Stafford00}]. Individual
predicted fits and observed values are shown in Figure \ref{fit_ind} for
six patients. We studied the distribution of residuals and the
agreement between predictions
and observations for plasma HIV RNA and CD4 count in Appendix \ref{apenB}.

\section{Discussion}\label{s5}

In this paper we present a method for estimating parameters of random effects
models based on differential equations when the origin of time is
unknown, taking also into account unbalanced data and
left-censored observations. This method was applied to a cohort of HIV patients
during primary infection using repeated measurements of two
markers: plasma HIV RNA and CD4$+$ T cell counts. The model gave reasonable
fits although the kinetics of the markers was very complex due to
nonlinear interaction between the virus and the target cells. Thanks to
the population approach, we
were able to describe the dynamics of markers during primary infection
using data from several hundred patients.

Predictions were obviously driven by the structure of the ODE system
that is based on biological knowledge.
Compared to a descriptive model, this mechanistic approach brings
additional information.
Typically, solutions of ODE systems led to oscillatory trajectories
[\citep{volterra1926fluctuations}]: these oscillations are dampened
as time passes so that the trajectory gets closer to the steady state
value. Observations
are partly consistent with the oscillating behavior. The peak of viral
load, the reality of which is
generally admitted, is nothing but the first oscillation. It can be
noted that the oscillations are generally more
dampened for more complex systems [\citep{Burg2009}].
In conclusion, the oscillatory trajectories produced by our
model are for a part the expression of a real phenomenon.

Predicted viral and CD4 setpoints (defined as
markers values at equilibrium) were quite similar to those that could
be observed.
Interestingly, the estimated value of infected cells (that are not
observed) was in
agreement with previous studies reporting a low concentration of
productive infected cells
[\citep{chun1997qlt}].

Because the approach was based on a mechanistic model, the parameters
have biological meanings.
For instance, the mean half-life of infected cells was estimated at
0.40 day. Previous estimations of this parameter were mainly performed
in patients treated by
antiretroviral therapy. These estimations varied from 0.7 day to 1.8
day according to the type of
treatment and the type of model used [\citep{Perelson96}; \citep{faulkner2003ppe}; \citep{huang2006bayesian}; \citep{huang2007modeling}].
Similarly, it is interesting to note that our estimate of virus
production (428 virions/cell/day)
is of the same order as what has been reported [\citep{dimitrov1993qhi}; \citep{levy1996pvl}; \citep{haase1999pbh}].
However, the biological relevance of the present model could still
be questioned with regards to other processes that have been ignored.
For instance, the cytotoxic T-cell
response plays a role in the control of the infection and the decrease
of the viral load by
killing the infected cells [\citep{Stafford00}], although the efficiency
of this
immunological response is debated [\citep{asquith2007vct}]. Moreover,
our estimate of target CD4 cells
half-life is a mixture of the half-life of quiescent cells and that of
activated cells [\citep{Mclean95}; \citep{vrisekoop2008spb}].

The uncertainty on the date of infection was taken into account to
improve the accuracy of the
estimates as compared to performing simple imputations. In addition,
we were able to derive the individual posterior distribution of
the date of infection according to the model. When enough information
is available the date
of infection can be estimated with good accuracy (Figure \ref{dist_ind}).
The sensitivity of the estimates to the assumptions about the priors
[equation (\ref{density}) in Section \ref{s22}]
was roughly proportional to the amount of
data available as illustrated by the posterior infection time
distribution of patient 41 compared
to patient 66 (Figure \ref{dist_ind}). We also performed a sensitivity
analysis which demonstrated a
certain robustness of the method. The lower bound of the infection time
distribution could be
better defined by using additional information such as antibody
dynamics. Unfortunately, this
information was not available in the present data set.

The main limitation for the use of dynamical models
is the parameter identifiability that led us to fix the value of
$\gamma$ (infection rate) and
$\mu_V$ (viral clearance). The clearance was fixed to 20 day$^{-1}$
according to recent studies with highly
repeated measures of HIV RNA after initiation of antiretroviral
treatment [\citep{perelson1997dch}; \citep{Ramratnam99}].
The chosen value of $\mu_V$ may influence the estimation of other
parameters such as $\pi$, as the viral
setpoint is essentially determined by the ratio $\frac{\pi}{\mu_V}$
[\citep{nowak1997avd}].
Identifiability may be improved by measuring
more compartments such as infected cells or by increasing the number of
repeated measurements [\citep{guedj2007pih}]. This is an important point
to consider in future studies, as the
issue of identifiability precludes the comparison with more complex models.

Finally, this method can be applied in other areas where either
the model is simpler or the amount of measured information greater,
so that identifiability is less an issue.

\begin{table}
\tabcolsep=3pt
\caption{Notation used in the manuscript}\label{t4}
\begin{tabular*}{\textwidth}{@{\extracolsep{\fill}}ll@{} }
\hline
\textbf{Notation} &  \textbf{Meaning} \\
\hline
$i$ $(1,\ldots,n)$&Subject \\
$X^i(t)$ &Vector of the $K$ components of the model at time $t$\\
$\xi^i$ &Vector of $p$ individual natural parameters \\
$\tilde\xi^i=\Psi(\xi^i)$ &Vector of $p$ individual
transformed parameters \\
$\phi_l$ &Intercept of the $l$th parameter \\
$z^i_l$ &Vectors of explanatory variables associated to the fixed
effects of the $l$th\\
& biological parameter \\
$\beta_l^i$ &Vectors of regression coefficients associated to the
fixed effects\\
$b^i=A u^i$&Individual vector of random effects of dimension $q$ \\
$A=(a_{l''l'})_{l'\leq l'' \leq q}$ &Lower triangular matrix with
positive diagonal elements $AA'=\Sigma$ \\
&(Cholesky decomposition)\\
$Y^i_{jm}$ &$j$th measurement of the $m$th compartment of subject
$i$\\
$d^i_{jm}$ &Date of measurement of $Y^i_{jm}$\\
$\tau^i$ &Date of infection of subject $i$ \\
$t^i_{jm}$ &Time between $\tau^i$ and $d^i_{jm}$\\
$g_m(\cdot)$ $(m=1,\ldots,M)$ &Nonlinear function for the $m$th observed
compartment\\
$\varepsilon^i_{jm}$ &Gaussian measurement errors with zero mean and
variances $\sigma_{m} ^{2}$\\
$d^i_-$ &Date of last negative HIV test of subject $i$\\
$d^i_+$ &Date of first positive HIV test of subject $i$\\
$f_{\tau^i}$ &Density of infection date $\tau^i$\\
$[\tau_{\inf}^i,\tau_{\sup}^i]$ &Support of the density $f_{\tau
^i}$\\
$\zeta^i_j$ &Detection limit of the $j$th measurement of subject
$i$\\
\hline
\end{tabular*}
\end{table}

\begin{appendix}\label{app}

\section{Score test for covariance of random effects}\label{apenA}
In \citep{Drylewicz2010scoretest}, we have developed score tests for
explanatory variables and variance of random effects in complex models.
We propose here
to develop a test for the covariance parameter of random effects. We
assume that the date of infection is known and introduce notation for a
conventional nonlinear model. For subject $i$ with $i=1,\ldots,n$, we
consider a model which specifies the distribution
of the observed vector ${\mathbf{Y}}^i=(Y^i_{j},j=1,\ldots,n^i)$:
\[
Y^i_{j}= g(t^i_{j},\tilde{\bolds{\xi}}^i) + \varepsilon^i_{j}, \qquad j=1,\ldots,n^{i},
\]
where $Y^i_{j}$ is the $j$th measurement for subject $i$ at the time
$t^i_{j}$, and the $\varepsilon^i_{j}$ are
independent Gaussian measurement errors with zero mean and variances
$\sigma^{2}$. The function $g(\cdot)$ is a twice
differentiable (generally nonlinear) function. The individual
parameters $\tilde{\bolds{\xi}}^i=(\tilde{\xi_l}^i,l=1,\ldots,p)$
are modeled as a linear form:
$\tilde\xi_l^i=\phi_l + b^i_l + z_l^i \bbeta_l$, where $\phi_{l}$
is the intercept,
and $z^i_l$ is the vector of explanatory variables associated to the
fixed effects of the $l$th parameter.
The $\bbeta_l$'s are vectors of regression coefficients.
We assume $b^i \sim\mathcal{N}(0,\Sigma)$, where $b^i$ is the individual
vector of random effects of dimension $q$. Let $A=(a_{l''l'})_{l'\leq
l'' \leq q}$ be the lower triangular matrix with positive diagonal
elements such that
$AA'=\Sigma$ (Cholesky decomposition). We can write $b^i=A \bu^i$
with $\bu^i \sim\mathcal{N}(\mathbf{0},\mathbf{I}_q)$. We denote by $\btheta=(\phi
_l,A,\beta_l,l=1,\ldots,p)$
the set of parameters of the model and by $\mathcal{L}^{\theta}_{\Y
^i|u^i}(Y^i|{\mathbf{u}})$ the likelihood of observations for subject $i$
given that the random effects $\ui$ take the value $\mathbf{u}$. Given
$\ui$, the $Y^i_{j}$ are independent,
so $\mathcal{L}^{\theta}_{\Y^i|u^i}(Y^i|{\mathbf{u}})= \prod_{j}\mathcal{L}^{\theta}_{\Y^i_{j}|u^i}(Y^i_{j}|{\mathbf{u}})$,
where
\[
\mathcal{L}^{\theta}_{\Y^i_{j}|u^i}(Y^i_{j}|{\mathbf{u}})=
\frac{1}{\sqrt{2\pi}\sigma}\exp\biggl\{-\frac{(Y^i_{j}-g(t^i_{j},\tilde
{\bolds{\xi}}^i))^2}{2\sigma^2}\biggr\}.
\]
The observed log-likelihood for subject $i$ is
\[
 L^{\theta}_i =\log\int_{\mathbb{R}^{q}}
\mathcal{L}^{\theta}_{\Y^i|u^i}(Y^i|{\mathbf{u}}) \phi(\mathbf{u})\, d\mathbf{u},
\]
where $\phi$ is the
multivariate normal density of $\mathcal{N}({\mathbf{0}},{\mathbf{I}}_{q})$.
We denote $L=L^{\theta}_1+\cdots+L^{\theta}_n$,
the global log-likelihood.

For simplicity and for illustrating our specific case, we assume that
only two random effects are included in the model ($u_l^i$ and
$u_{l'}^i$). We want to test whether
$b_l^i$ and $b_{l'}^i$ are independent. The null hypothesis $H_0$ is
``$a_{ll'}=0$''. The scores $U_{ll'}^{i}$ (individual score for
$a_{ll'}$) are obtained
integrating $U_{ll'|u^i}^{i}$ using Louis' formula [\citep{Louis82}]:
\[
U_{ll'}^{i}= (\mathcal{L}_{\mathcal{O}^i}) ^{-1} \int_{\mathbb
{R}^{2}} \mathcal{L}^{\theta}_{\Y^i|u^i}({\mathbf{u}})
U_{ll'|u^i}^i({\mathbf{u}}) \phi({\mathbf{u}}) \,d{\mathbf{u}},
\]
where
\[
U_{ll'|u^i}^{i}(\mathbf{u})= \sum_{j=1,n^i} \frac{1}{\sigma^{2}} \bigl(
Y^i_{j}- g(X(t^i_{j}, \xiit)) \bigr) \biggl( u_{l'}^i a_{ll}
\frac{\partial g(X(t^i_{j}, \xiit))}{\partial\xii_l } \biggr) .
\]
Under the null hypothesis, we propose the following statistic:
\[
S=\frac{U^{\sullet}_{ll'|H_0}}{\sqrt{\widehat\var U^{\sullet
}_{ll'|H_0}}},
\]
where $U^{\sullet}_{ll' |H_0}$ is the sum of individual scores
$U_{ll'}^{i}$ calculated under $H_0$ and
$\widehat\var\,{U^{\sullet}_{ll'|H_0}}$ is a consistent estimator of
$\var{U^{\sullet}_{ll'|H_0}}$. $S$ has an asymptotic standard normal
distribution under $H_0$.
We take $\widehat\var\,{U^{\sullet}_{ll'|H_0}}= \sum
_{i=1,n}{U^{i2}_{ll'|H_0}}$.

\section{Goodness of fit}\label{apenB}

\begin{figure}

\includegraphics{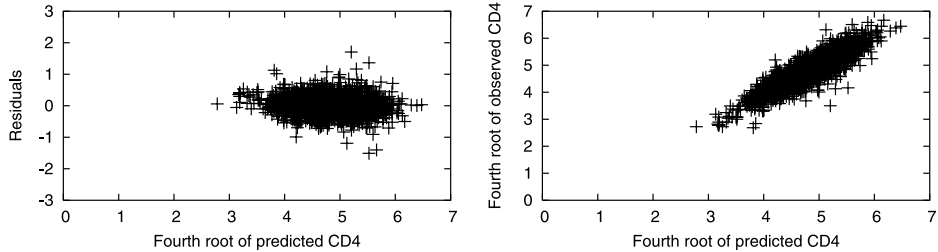}

\caption{Residuals of fourth-root of CD4 count with respect to
predictions (\textup{left}) and observations with respect to predictions
(\textup{right}).}\label{f5}
\end{figure}

\begin{figure}

\includegraphics{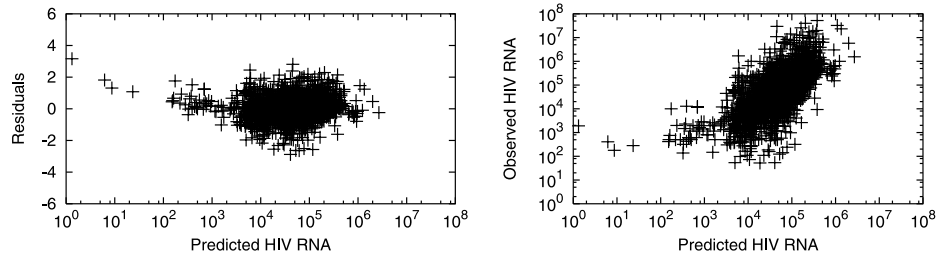}

\caption{Residuals of HIV RNA with respect to predictions (\textup{left}) and
observations with respect to predictions (\textup{right}), the left-censored
measurements of HIV RNA have been excluded (72 among 3038
measurements).}\label{f6}
\end{figure}

Figures \ref{f5} and \ref{f6} present the residuals with respect to the
predictions and the observations with respect to the predictions for
the fourth root of CD4 count and the base 10-logarithm of HIV RNA load.
The residuals are computed using empirical bayes and predictions from
the most probable date of infection for each patient. We excluded
left-censored HIV RNA measurements (72 among 3038 measurements).

The residuals are well distributed around 0 and there is a good
agreement between predictions and observations for both HIV RNA and CD4
count. The four outliers observed on the left of HIV RNA figures are
due to patients having only one detectable measurement of HIV RNA
during their follow-up and the only detectable value was low ($<$$3 \log
_{10}$ copies/mL).

\end{appendix}

\section*{Acknowledgment}

The authors would like to thank S. Walker for suggestions and criticisms
concerning the manuscript.

%

\section*{CASCADE collaboration}

\textit{Steering Committee}: Julia Del Amo (Chair), Laurence Meyer
(Vice Chair),
Heiner Bucher, Genevi\`{e}ve Ch\^{e}ne, Deenan Pillay, Maria Prins,
Magda Rosinska, Caroline Sabin, Giota Touloumi.

\textit{Co-ordinating Centre}: Kholoud Porter (Project Leader), Sara Lodi,
Sarah Walker, Abdel Babiker, Janet Darbyshire.

\textit{Clinical Advisory Board}: Heiner Bucher, Andrea de Luca,
Martin Fisher,
Roberto Muga.

\textit{Collaborators}: Australia Sydney AIDS Prospective Study and
Sydney Primary
HIV Infection cohort (John Kaldor, Tony Kelleher, Tim Ramacciotti, Linda
Gelgor, David Cooper, Don Smith); Canada South Alberta clinic (John Gill);
Denmark Copenhagen HIV Seroconverter Cohort (Louise Bruun Jorgensen,
Claus Nielsen, Court Pedersen); Estonia Tartu Ulikool (Irja Lutsar);
France Aquitaine cohort (Genevi\`{e}ve Ch\^{e}ne, Francois Dabis, Rodolphe
Thi\'{e}baut, Bernard Masquelier), French Hospital Database (Dominique
Costagliola, Marguerite Guiguet), Lyon Primary Infection cohort
(Philippe Vanhems), SEROCO cohort (Laurence Meyer, Faroudy Boufassa);
Germany German cohort (Osamah Hamouda, Claudia Kucherer); Greece Greek
Haemophilia cohort (Giota Touloumi, Nikos Pantazis, Angelos Hatzakis,
Dimitrios Paraskevis, Anastasia Karafoulidou); Italy Italian Seroconversion
Study (Giovanni Rezza, Maria Dorrucci, Benedetta Longo, Claudia Balotta);
Netherlands Amsterdam Cohort Studies among homosexual men and drug users
(Maria Prins, Liselotte van Asten, Akke van der Bij, Ronald Geskus,
Roel Coutinho);
Norway Oslo and Ulleval Hospital cohorts (Mette Sannes, Oddbjorn
Brubakk, Anne Eskild,
Johan N. Bruun); Poland National Institute of Hygiene (Magdalena
Rosinska); Portugal
Universidade Nova de Lisboa (Ricardo Camacho); Russia Pasteur Institute
(Tatyana Smolskaya); Spain Badalona IDU hospital cohort (Roberto Muga),
Barcelona IDU Cohort (Patricia Garcia de Olalla), Madrid cohort
(Julia Del Amo, Jorge del Romero), Valencia IDU cohort (Santiago P\'
{e}rez-Hoyos,
Ildefonso Hernandez Aguado); Switzerland Swiss HIV cohort (Heiner
Bucher, Martin
Rickenbach, Patrick Francioli); Ukraine Perinatal Prevention of AIDS Initiative
(Ruslan Malyuta); United Kingdom Edinburgh Hospital cohort (Ray
Brettle), Health
Protection Agency (Valerie Delpech, Sam Lattimore, Gary Murphy, John
Parry, Noel Gill),
Royal Free haemophilia cohort (Caroline Sabin, Christine Lee), UK
Register of HIV
Seroconverters (Kholoud Porter, Anne Johnson, Andrew Phillips, Abdel
Babiker, Janet
Darbyshire, Valerie Delpech), University College London (Deenan
Pillay), University
of Oxford (Harold Jaffe).

\printaddresses

\end{document}